\documentclass{osa-article}

\journal{oe}

\begin{document}

\title{Exploring higher Jaynes-Cummings doublet in cavity quantum electrodynamics system with a broadband squeezed vacuum injection}

\author{D. Q. Bao,\authormark{1} C. J. Zhu,\authormark{1,*} Y. P. Yang\authormark{1,$\dag$} and G. S. Agarwal\authormark{2,$\dag\dag$}}

\address{\authormark{1}MOE Key Laboratory of Advanced Micro-Structured Materials, School of Physics Science and Engineering, Tongji University, Shanghai 200092, China\\
\authormark{2}Institute for Quantum  Science and Engineering, and Departments of Biological and Agricultural Engineering, Physics and Astronomy, Texas A\&M University, College Station, TX 77843, USA}

\email{\authormark{*}cjzhu@tongji.edu.cn} 
\email{\authormark{$\dag$}yang\_yaping@tongji.edu.cn}
\email{\authormark{$\dag\dag$}girish.agarwal@tamu.edu}



\begin{abstract}
We investigate the cavity excitation spectrum and the photon number distribution in a cavity QED system driven by a broadband squeezed vacuum. In an empty cavity, we show that only states with even number of photons can be measured under resonant condition since the squeezed vacuum consists of states with even number of photons only. When a single atom is trapped in the cavity, the strong coupling between the atom and cavity results in energy splittings of the system, and there exist two peaks in the cavity excitation spectrum at two-photon transition frequencies. At the central frequency, however, all photon states can be detected because of the interaction between the atom and cavity. Therefore, it can be used to detect whether a single atom is trapped in the cavity. We also show that the squeezed vacuum can promote multiphoton excitations in the cavity. Using a coherent probe field, it is possible to explore higher Jaynes-Cummings doublet even if the probe field intensity is very weak. 
\end{abstract}

\section{Introduction}
Nonlinearity induced quantum optical effects in cavity quantum electrodynamics (QED) systems have been intensively studied in past few decades, including the  squeezed excitation~\cite{Turchette1998}, multiphoton excitation~\cite{Fink2008,Bishop2009,Fink2009}, photon blockade phenomenon~\cite{Birnbaum2005,Hamsen2017,Zhu2017} and so on~\cite{agarwal2016perfect,wei2018coherent}. Many of these properties have been explained theoretically and observed experimentally~\cite{Scully1997,Agarwal2013}. The investigations on nonlinear features of optical field not only reveal some fundamental questions about cavity QED, but also lead to many practical applications in many fields, including high precision measurements~\cite{Yamamoto1986}, optical communications~\cite{Hirota} and optical information processing~\cite{Andersen2015}. 

In general, a high-finesse cavity is required to produce nonlinear effects because strong coupling between a quantum emitter and cavity can be achieved. In this case, dynamical evolution of the quantum emitter in strong coupling regime is intimately tied with the resonant or non-resonant modes in the cavity. More specifically, the quantum emitter can be detected in a desired state if the cavity is well prepared~\cite{ding2017cross}. With the development of current experimental technologies, strong coupling between quanta and cavity becomes possible not only in traditional cavity QED systems but also in circuit QED systems. 
Particularly, many nonlinear features such as two-photon and three-photon excitations have been observed experimentally in the circuit QED systems because extremely strong coupling  can be realized by the large dipole coupling strength and the long coherence time of artificial ``atoms''~\cite{Niemczyk2010,Yoshihara2017,Forn-Diaz2017}. 

To explore much higher Jaynes-Cummings doublet states, we make the cavity driven by a coherent probe field and a broadband squeezed vacuum simultaneously. Physically, the squeezed vacuum will result in excitations of state with even photon numbers, which provides a possibility to climb much higher Jaynes-Cummings doublet even if the probe field is very weak. Currently, the squeezed light can be successfully generated by the parametric oscillator or parametric down converter~\cite{wu1986generation}. Due to its potential applications in the fields of quantum measurement, optical communication, and quantum information processing, squeezed vacuum state has been extensively studied~\cite{furusawa1998unconditional,tan2005continuous,honda2008storage,haus2012electromagnetic,Kono2017}.

In this paper, we carefully study the cavity excitation spectrum and the photon number distributions in a single atom-cavity QED system. We show many interesting features arising from the squeezed vacuum by comparing with the case of thermal field injection. In an empty cavity, for example, we show that the properties of the squeezed vacuum can be observed by detecting the photon number distribution. In a single-atom-cavity QED system, however, we find that the properties of the squeezed vacuum are changed significantly due to the strong coupling between the atom and cavity, which are reflected in both the cavity excitation spectrum and the photon number distribution. Based on these features, it is possible to detect whether a single atom is trapped in the cavity. We also show that the squeezed vacuum boosts the multiphoton excitations and can be used to explore higher Jaynes-Cummings doublet states with a weak coherent field.

\section{System model}
To begin with, we consider that a single two-level atom with resonant transition frequency $\omega_A$ is strongly coupled to a single-mode cavity with frequency $\omega_C$. 
\begin{figure}[h!]
\centering
\includegraphics[width=0.6\linewidth]{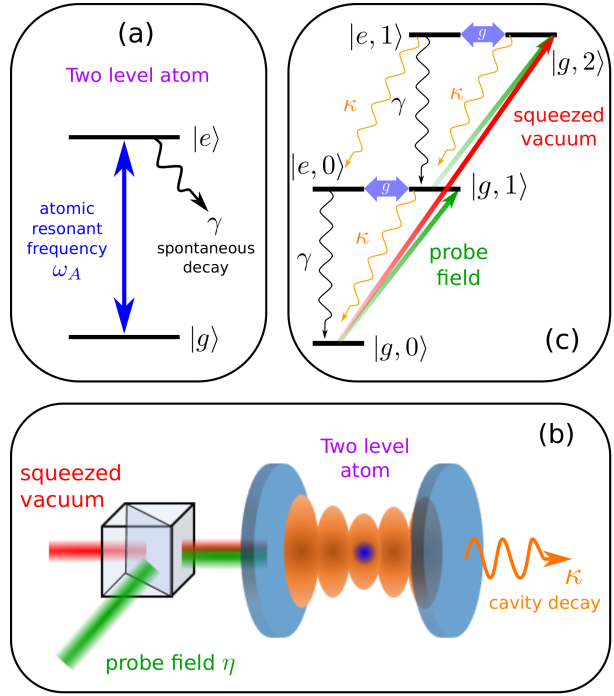}
\caption{The configuration and the dressed state picture of the system. A two-level atom [panel (a)] with resonant transition frequency $\omega_A$ is trapped at the antinode of an intracavity light field of frequency $\omega_C$ [panel (b)]. The ground (excited) state of the two level atom is labeled as $|g\rangle$ ($|e\rangle$). The cavity is coherently driven by a probe field of frequency $\omega_p$, and a broadband squeezed vacuum with central frequency $\omega_S$ at the same time. The Rabi frequency of the probe field is labeled as $\eta$. Here, $\kappa$ and $\gamma$ denote the spontaneous decay of the atom and the damping rate of the cavity, respectively. The energy-level structure and transition pathways of the system are shown in panel (c).}~\label{fig:fig1}
\end{figure}
As shown in Fig.~\ref{fig:fig1}(a), the ground (excited) state of this atom is labeled as $|g\rangle$ ($|e\rangle$). The cavity is driven by a \underline{broadband} squeezed vacuum with central frequency $\omega_S$, and a coherent probe field with  angular frequency $\omega_p$ simultaneously [see panel (b)]. Here, we assume $\omega_C=\omega_A=\omega_0$ and $\omega_p=\omega_S$ for simplicity.

In general, the atom-cavity strongly-coupled system can be well described by the Jaynes-Cummings Hamiltonian in a frame rotating at the frequency of the coherent probe field frequency $\omega_S$, i.e., 
\begin{eqnarray}
H_{\rm I}&=&\hbar\left[\Delta_C a^\dagger a+\Delta_A\sigma_{+}\sigma_{-}+g(a^\dagger\sigma_{-}+a\sigma_{+})+\eta(a+a^\dagger)\right],
\end{eqnarray}
where $\Delta_C=\Delta_A=\omega_0-\omega_S$ and $g$ is the coupling strength between the atom and cavity field and $\eta$ is the Rabi frequency of the probe field. $\sigma_{+}=|e\rangle\langle g|$ ($\sigma_{-}=|g\rangle\langle e|$) denotes the raising (lowering) operator of the two level atom, and $a^\dagger$ ($a$) denotes the cavity photon creation (annihilation) operator. The effect of the squeezed bath is not contained in Eq.~(1) but would be included in the master equation.

In the normalized unit with $\hbar=1$, the evolution of the density matrix $\rho$ of this system is then given by 
\begin{eqnarray}\label{eq:master}
\frac{d\rho}{dt}&=&-i[H_{\rm I},\rho]+{\cal L}_{\rm atom}\rho+{\cal L}_{\rm cav}\rho 
\end{eqnarray}
where ${\cal L}_{\rm atom}\rho=-\gamma(\sigma_{+}\sigma_{-}\rho-2\sigma_{-}\rho\sigma_{+}+\rho\sigma_{+}\sigma_{-})$ originates from the spontaneous decay of this two-level atom at rate $\gamma$. The effect of the injected broadband squeezed vacuum is given by the last term in Eq.~(2) and can be expressed in the form of~\cite{note2}
\begin{eqnarray}
{\cal L}_{\rm cav}\rho &=& -\kappa(1+N)(a^\dagger a\rho-2a\rho a^\dagger+\rho a^\dagger a)-\kappa N(aa^\dagger\rho-2a^\dagger\rho a+\rho aa^\dagger)\nonumber\\
& & -\kappa M (a^\dagger a^\dagger\rho-2a^\dagger\rho a^\dagger+\rho a^\dagger a^\dagger)-\kappa M^\ast(aa\rho-2a\rho a+\rho aa),
\end{eqnarray}
where 
$N=\sinh^2{(r)}$ and $M=\cosh{(r)}\sinh{(r)}{\rm e}^{i\phi}$ are the photon number and the strength of two-photon correlation in the injected squeezed vacuum, respectively. Here, $r$ ($\phi$) is the squeezing parameter (phase). Note that if squeezed vacuum is replaced by the usual vacuum, (i.e., if $r=0$), then the phase dependent terms ($M$ terms) drop out and only the very first term in Eq.~(3) survives. We also note that the limit of thermal reservoir can be obtained formally by setting $M=0$ and $N=\bar{n}$ where $\bar{n}$ is the number of thermal photons at cavity frequency.  

\section{Empty cavity system}
Before studying such a complicated system, we first review the case of an empty cavity driven by a thermal field or a broadband squeezed vacuum, respectively (i.e., setting $g=0$ and $\eta=0$). As we all known, if the cavity is driven by a thermal field, i.e., $N=\bar{n}\neq0$ but $M=0$, the average photon number in the cavity $\langle a^\dagger a\rangle_{\rm empty}^{\rm th}=N=\bar{n}$, which is independent of the thermal field frequency. Likewise, if the cavity is driven by a squeezed vacuum (i.e., $N=\sinh^2{(r)}$ and $M=\cosh{(r)}\sinh{(r)}{\rm e}^{i\phi}$), we can also analytically calculate the average photon number in the cavity, which is given by $\langle a^\dagger a\rangle_{\rm empty}^{\rm sq}=N=\sinh^2{(r)}$. 
%
\begin{figure}[h!]
\centering
	\includegraphics[width=0.8\linewidth]{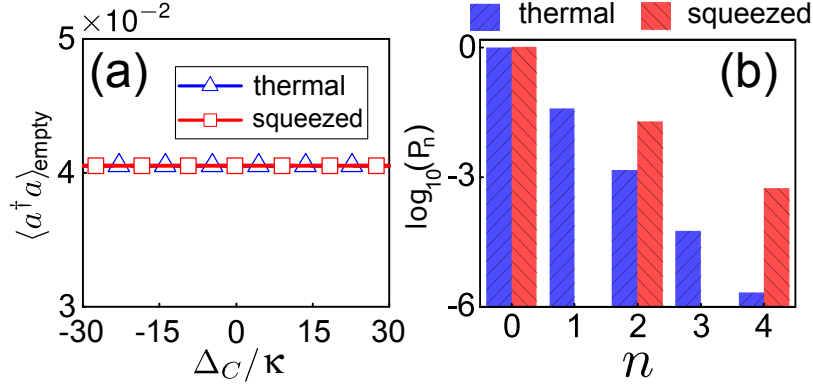}
	\caption{(Color online) (a) The cavity excitation spectrum versus the normalized detuning $\Delta_C/\kappa$ for an empty cavity ($g=0$) driven by a thermal field (blue line with triangles) and a squeezed vacuum (red line with squares), respectively. (b) The photon number distribution at cavity resonant frequency, i.e., $\Delta_C=0$. The blue bars indicate the thermal field injection, but the red ones indicate the squeezed vacuum injection. The system parameters are given by $r=0.2$ and $\phi=\eta=0$.}
~\label{fig:fig2}
\end{figure}
In Fig.~\ref{fig:fig2}(a), we show that numerical results of the average photon number, where the red line with squares indicates the squeezed vacuum driving, but the blue line with triangles indicates the thermal field driving. The system parameters are chosen as $r=0.2$ and $\phi=0$. In the case of thermal driving, we take $N=\bar{n}=\sinh^2{(r)}$ and $M=0$ for mathematical simplicity. In order to have a fair comparison of thermal and squeezed bath cases, we set the photon number equal. 

Although they have the same characteristics in the cavity excitation spectrums, thermal/squeezed vacuum driving causes different properties in the photon distributions. In Fig.~\ref{fig:fig2}(b), we show the photon distribution $P_n$ when the driving field is resonant to the cavity, i.e., $\Delta_C=0$. In the squeezed vacuum driving case, the probability of the photon number distribution $P_{2k+1}=0$ for all states with odd number of photons, and $P_{2k}=(2k)!\tanh^{2k}{(r)}/[2^{2k}(k)!\cosh{(r)}]$ for all states with even number of photons [see Fig.~\ref{fig:fig2}(b), red bars]. In the thermal driving case, however, the probability of the photon number distribution is given by $P_{n}=N^n/(N+1)^{n+1}$ [see the blue bars] (i.e., all states have populations). Only when the frequency of the squeezed driving field is tuned far off-resonant to the cavity frequency, the photon distributions are the same as that of the thermal driving case [More details are discussed in the appendix].

\section{Single atom cavity QED system}
\begin{figure}[h!]
\centering
	\includegraphics[width=0.8\linewidth]{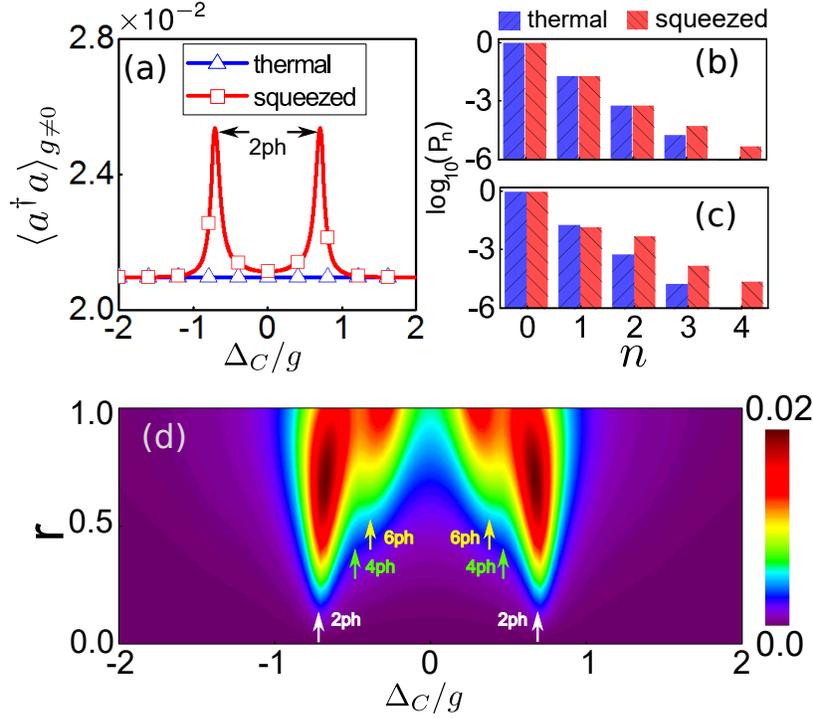}
	\caption{(Color online) Panel (a) shows the average photon number in a single atom-cavity QED system driven by a thermal field (blue curve with triangles) and a squeezed vacuum (red curve with squares), respectively. Panel (b) and (c) show the corresponding photon number distribution in the cavity resonant frequency ($\Delta_C=0$) and the two-photon resonant frequency ($\Delta_C=\sqrt{2}g/2$) of the driving field, respectively. The blue bars indicate the results for thermal field driving, but the red bars indicate the results for squeezed vacuum driving. Here, we choose $g/\gamma=15$, $r=0.2$ and other system parameters are the same as those used in Fig.~\ref{fig:fig2}. Panel (d) shows the change in the average photon number $\Delta n=\langle a^\dagger a\rangle_{g\neq0}-\langle a^\dagger a\rangle_{g\neq0}^{\rm th}$ versus the normalized detuning $\Delta_C/g$ and the parameter $r$, respectively. The detunings $\Delta_c/g=\pm\sqrt{n}/n$ stand for the $n$-photon excitations.}
	~\label{fig:fig3}
\end{figure}

Now, we consider the case that a two-level atom is trapped in the cavity (i.e., $g\neq0$), where the master equation [Eq.~(\ref{eq:master})] can only be solved numerically in the strong coupling regime since multiphoton transition pathways must be taken into account. In Fig.~\ref{fig:fig3}(a), we plot the cavity excitation spectrums for the thermal field driving (blue curve with triangles) and the squeezed vacuum driving (red curve with squares) as a function of the normalized cavity detuning $\Delta_C/g$. In the case of the thermal field driving, the cavity excitation spectrum (blue curve) is similar to the case of empty cavity, which is flat and independent of the driving field frequency,  but its magnitude decreases slightly due to the atomic damping. In the case of the squeezed filed driving, however, there exist two peaks in the cavity excitation spectrum at the two-photon excitation frequencies, i.e., $\Delta_C=\pm\sqrt{2}g/2$ (red curve), corresponding to $|g,0\rangle\rightarrow|g,2\rangle$ transitions as depicted in Fig.~\ref{fig:fig1}(c). It is worthy to point out that the frequency shift is caused from the strong coupling. Here, we choose the coupling strength between the atom and cavity is $g/\gamma=15$, and other system parameters are the same as those used in Fig.~\ref{fig:fig2}. When the frequency of the squeezed vacuum is far away from the two-photon resonant frequency, it is found that the average photon number $\langle a^\dagger a\rangle_{g\neq0}^{\rm sq}=\langle a^\dagger a\rangle_{g\neq0}^{\rm th}$ because the properties of the squeezed vacuum become weak, and the thermal excitations are dominant.

%

To show properties of the squeezed vacuum more clearly, we plot the photon distributions of the output field at cavity resonant frequency [$\Delta_C=0$, panel (b)] and two-photon resonant frequency [$\Delta_C=\sqrt{2}g/2$, panel (c)], respectively. Here, the blue bars indicate the thermal driving case, while the red bars indicate the squeezed vacuum driving case. As shown in panel (b), the photon distributions for the squeezed vacuum driving are almost same as those for the thermal driving when the driving field frequency is equal to the cavity frequency. This is because the squeezed vacuum is off-resonant to the cavity due to the energy splitting caused by the strong coupling between the atom and cavity. When the squeezed vacuum frequency is tuned to the two-photon resonant frequency, the properties of the squeezed vacuum become strong enough to be observed. As shown in Fig.~\ref{fig:fig3}(c), the amplitudes of the photon distributions for squeezed vacuum driving are much larger than those for thermal driving if the photon number of Fock state $n\geq2$, which provides a possibility to observe multiphoton transitions and explore high-order nonlinearity in the cavity QED system. 

In Fig.~\ref{fig:fig3}(d), we plot the fluctuations of the average photon number $\Delta n=\langle a^\dagger a\rangle_{g\neq0}-\langle a^\dagger a\rangle_{g\neq0}^{\rm th}$ as functions of the normalized detuning $\Delta_C/g$ and the squeezing parameter $r$. Obviously, only two peaks at two-photon resonant frequencies, i.e., $\Delta_C=\pm\sqrt{2}g/2$ can be observed for weak squeezing strength (i.e., small $r$) since high-order transition processes are too weak to be excited. However, with the squeezing strength $r$ increasing, four-photon excitations ($\Delta_C=\pm g/2$) and even six-photon excitations ($\Delta_C=\pm\sqrt{6}g/6$) become strong enough to be detected. As a result, there exist many peaks in the excitation spectrum, corresponding to the multiphoton excitation processes that are hard to be observed by only using a thermal field.

Based on these interesting properties exhibited above, we propose a new method to detect whether an atom is trapped in the cavity by using squeezed vacuum. In this proposal, we assume that there exists a single atom in the cavity, and the squeezed vacuum frequency is the same as the cavity frequency, i.e., $\Delta_C=0$. As a result, one can detect the existence of a single atom by measuring the probability of one-photon state distribution, i.e., $P_1$. If $P_1>0$ is measured, a single atom is successfully trapped in the cavity. Otherwise, $P_1=0$ can be measured due to the property of the squeezed vacuum. 
\begin{figure}[h!]
\centering
	\includegraphics[width=0.8\linewidth]{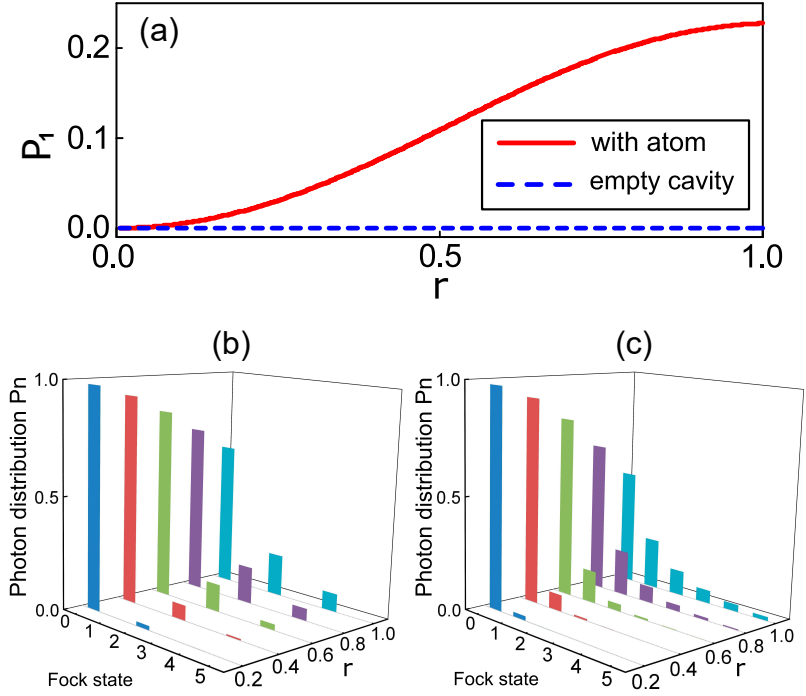}
	\caption{(Color online) (a) The population of one-photon state is plotted as a function of the squeezing strength $r$ in an empty cavity (blue dashed line) and a single-atom-cavity QED system (red solid curve), respectively. Panels (b) and (c) demonstrate the photon number distribution with different parameter $r$ in an empty cavity and a single-atom-cavity QED system driven by a squeezed vacuum, respectively. The detuning is chosen as $\Delta_C=0$ and other system parameters are the same as those used in Fig.~\ref{fig:fig3}.}
	~\label{fig:fig4}
\end{figure}
As shown in Fig.~\ref{fig:fig4}(a), the stronger the squeezed vacuum is, the higher the probability of one-photon state distribution is. This excitation as shown in Fig.~1(c) is via $|g,2\rangle\overset{g}\rightarrow|e,1\rangle\overset{\kappa}\rightarrow|e,0\rangle$. In Fig.~\ref{fig:fig4}(b) and (c), we demonstrate the photon state  distribution in an empty cavity [panel (b)] and a single atom-cavity QED system [panel (c)], respectively, by varying the squeezing strength $r$. In an empty cavity, it is clear to see that only the states with even number of photons can be detected i.e., $P_{2k+1}=0$ and $P_{2k}\neq0$~\cite{Agarwal2013}. With the squeezing strength $r$ increasing, more and more states with even number of photons can be detected. However, in presence of the atom, the states with odd number of photons can also be detected as demonstrated in panel (c). This is because the cavity absorbs the squeezed vacuum via the two-photon process, and then couples to the atomic excited state via the interaction between the atom and cavity. Therefore, it is possible to measure photons with odd number via the cavity damping as shown in Fig.~\ref{fig:fig1}(c). It is also noticed that the larger the squeezing strength is, the more the states with odd number of photons can be measured. 

\section{Competing effects of the coherent and squeezed radiations}
\begin{figure}[h!]
\centering
	\includegraphics[width=0.8\textwidth]{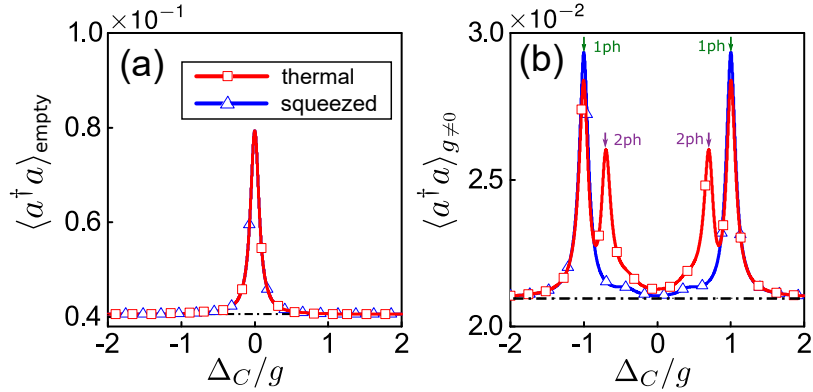}
	\caption{(Color online) Panels (a) and (b) show the average photon number as a function of the normalized detuning $\Delta_C$ in an empty cavity and a single-atom-cavity QED system, respectively. The blue (red) curves represent the case that the thermal (squeezed) field is injected into the system with a coherent field. Here, the coherent field is chosen as $\eta/\gamma=0.2$ for both panels (a) and (b). Other system parameters are the same as those used in Fig.~\ref{fig:fig3}. The black dash-dotted lines indicate the case that the cavity is only driven by a thermal field.}
	~\label{fig:fig5}
\end{figure}

In the following, we study the case that the cavity is driven by a coherent field $\eta$ and an incoherent field (squeezed or thermal field) simultaneously. In this case, the one-photon excitations by the coherent field and the multiphoton excitations by the incoherent field compete with each other, which results in many interesting interference phenomena. Firstly, we consider the case of an empty cavity driven by a coherent probe field and an incoherent field simultaneously. As shown in Fig.~\ref{fig:fig5}(a), the cavity excitation spectrum for the thermal field is the same as that for the squeezed vacuum, and there exists a single peak at $\Delta_C=0$ in the cavity excitation spectrum, which is caused by the coherent field. Here, we choose $\eta/\gamma=0.2$, $r=0.2$ and other system parameters are the same as those used in Fig.~\ref{fig:fig3}.

When an atom is trapped in the cavity, the cavity excitation spectrum exhibits many new features due to the competition between one-photon and two-photon excitations. For example, in the case of thermal field injection, there exist two peaks at $\Delta_C=\pm g$ in the cavity excitation spectrum, corresponding to the one-photon excitations [see the blue curve in panel (b)]. Compared with the result shown in Fig.~\ref{fig:fig3}(a), we find that these two peaks attribute to the coherent field. However, in the case of squeezed vacuum injection, there exist four peaks, corresponding to the one-photon excitations at $\Delta_C=\pm g$ and two-photon excitations at $\Delta_C=\pm\sqrt{2}g/2$, respectively. Obviously, the squeezed vacuum results in the two-photon excitations, whose amplitude is close to that of the one-photon excitation arising from the coherent field. Comparing these two cases, we find that high-order photon transition processes are much easier to be excited if the cavity is driven by the squeezed vacuum. 

%
\begin{figure}[h!]
\centering
	\includegraphics[width=0.8\textwidth]{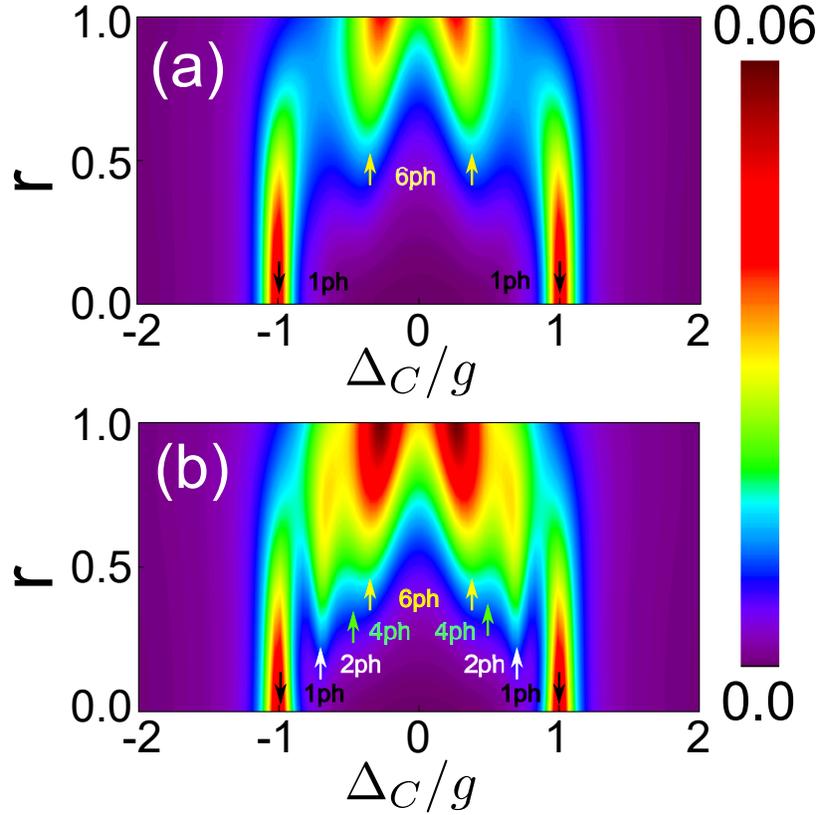}
	\caption{(Color online) The change in the average photon number $\Delta n=\langle a^\dagger a\rangle_{g\neq0}-\langle a^\dagger a\rangle_{g\neq0}^{\rm th}$ versus the normalized detuning $\Delta_C/g$ and the parameter $r$ respectively. Panel (a) corresponds to the thermal field injection, but panel (b) corresponds to the squeezed vacuum injection. Here, we choose $\eta/\gamma=0.5$ and other system parameters are the same as those used in Fig.~\ref{fig:fig3}. The coherent field is on in both panels (a) and (b). The detunings $\Delta_c/g=\pm\sqrt{n}/n$ stand for the $n$-photon excitations.}
	~\label{fig:fig6}
\end{figure}
We next bring the effects due to phase characteristics of the squeezed bath. In Fig.~\ref{fig:fig6}, we plot the change in the average photon number $\Delta n=\langle a^\dagger a\rangle_{g\neq0}-\langle a^\dagger a\rangle_{g\neq0}^{\rm th}$ by changing the parameter $r$. Here, we choose $\eta/\gamma=0.5$ to compete with the squeezed/thermal field. In the case of thermal field injection [see panel (a)], it is clear to see that only one-photon excitations can be observed for weak parameter $r$. If the thermal field driving strength $r>\eta$, all multiphoton excitations merge together. As the driving strength $r$ increases, the amplitudes of the multiphoton excitations induced by the thermal field become as strong as the one-photon excitations. In the case of squeezed vacuum injection, two-photon excitations induced by the squeezed vacuum compete with the one-photon excitations induced by the coherent field. As a result, one-photon and two-photon excitations can be clearly observed and well separated under even if $r<\eta$. When the squeezing strength $r$ is close to the coherent field $\eta$, four-photon and much higher-order excitations can be excited. As the parameter $r$ increases, high-order excitations are boosted by the squeezed vacuum, resulting in that the amplitudes of multiphoton excitations become much stronger than those of one-photon excitations [see panel (b)].

\section{Conclusion}
In summary, we have studied the cavity QED system driven by a squeezed vacuum, which exhibits many interesting features. In an empty cavity system, although the cavity excitation spectrum for the squeezed vacuum is the same as that for the thermal field, the photon number distributions are different at the cavity resonant frequency. 
When a two-level atom is trapped in the cavity, states with odd number of photons get populated due to the coupling between the atom and cavity, and two-photon excitations are dominant for weak squeezing strength. As the squeezing strength increases, multi photon excitations can also be observed, which results in many peaks in the cavity excitation spectrum. Based on these properties arising from the squeezed vacuum, we propose a method to detect whether a single atom is trapped in the cavity. We also show that the squeezed vacuum promotes the multiphoton excitations and can be used to climb higher Janyes-Cummings doublet with a weak coherent probe field.

\appendix
\section*{Appendix A: Photon distributions in an empty cavity}
\setcounter{equation}{0}
\renewcommand{\theequation}{A{\arabic{equation}}}
To show the photon state distributions, we evaluate the population in each photon state by solving the master equation. In the case of thermal field driving, we can obtain a set of equations for the elements of density matrix $p_{mn}=\langle m|\rho |n\rangle$, which is given by
\begin{align}
\frac{d}{dt}p_{n,n}=&-2\kappa(1+N)[np_{n,n}-(n+1)p_{n+1,n+1}]\nonumber\\
&-2\kappa N[(n+1)p_{n,n}-np_{n-1,n-1}]
\end{align}
It is well known that the steady-state solution of the above equation can be obtained easily, yielding $p_{nn}=N^n/(1+N)^{n+1}$~\cite{Agarwal2013}. We note that the photon distribution is independent of the detuning $\Delta_C$.

In the case of squeezing field driving, we can also obtain a set of equations for $p_{m,n}$, yielding
\begin{align}
\frac{d}{dt} p_{n,n}=&-2\kappa(1+N)[np_{n,n}-(n+1)p_{n+1,n+1}]-2\kappa N[(n+1)p_{n,n}-np_{n-1,n-1}]\nonumber\\
&-\kappa M(\sqrt{n(n-1)}p_{n-2,n}-2\sqrt{n(n+1)}p_{n-1,n+1}+\sqrt{(n+1)(n+2)}p_{n,n+2})\nonumber\\
&-\kappa M(\sqrt{n(n-1)}p_{n,n-2}-2\sqrt{n(n+1)}p_{n+1,n-1}+\sqrt{(n+1)(n+2)}p_{n+2,n})\\
\frac{d}{dt}p_{m,n}=&-i\Delta_C(m-n)p_{m,n}-\kappa(1+N)[(m+n)p_{m,n}-2\sqrt{(m+1)(n+1)}p_{m+1,n+1}]\nonumber\\
&-\kappa N[(m+n+2)p_{m,n}-2\sqrt{mn}p_{m-1,n-1}]-\kappa M(\sqrt{m(m-1)}p_{m-2,n}\nonumber\\
&-2\sqrt{m(n+1)}p_{m-1,n+1}+\sqrt{(n+1)(n+2)}p_{m,n+2})-\kappa M(\sqrt{(m+1)(m+2)}p_{m+2,n}\nonumber\\
&-2\sqrt{(m+1)n}p_{m+1,n-1}+\sqrt{n(n-1)}p_{m,n-2}).
\end{align}
To obtain analytical solutions we consider the case that the driving field is very weak (for example, $r<0.1$). As a result, we can assume $p_{00}\approx1$ and safely neglect all equations of $p_{n,n}$ for $n\geq3$. Then, under the steady-state approximation, Eqs. (4) and (5) can be reduced to 
\begin{align}
&(1+3N)p_{11}-2(1+N)p_{22}-2\sqrt{2}M{\rm Re}(p_{20})=N,\\
&2Np_{11}-(2+5N)p_{22}-\sqrt{2}M{\rm Re}(p_{20})=0,\\
&\kappa M(1-2p_{11}+p_{22})+\sqrt{2}[\kappa(3N+1)+i\Delta_C]p_{20}=0.
\end{align}

Solving Eq. (8), we have $p_{20}=-\kappa M(1-2p_{11}+p_{22})/\sqrt{2}[\kappa(3N+1)+2i\Delta_C]$. Inserting into Eqs. (6) and (7) we can obtain the populations in one-photon and two-photon states, respectively, which reads
\begin{align}
p_{11}=&\frac{W(2+2N+W)-(N+W)(4+10N+W)}{(4N+2W)(2+2N+W)-(1+3N+2W)(4+10N+W)},\\
p_{22}=&\frac{(1+3N+2W)W-(4N+2W)(N+W)}{(4N+2W)(2+2N+W)-(1+3N+2W)(4+10N+W)},
\end{align}
where $W=-2\kappa^2M^2(3N+1)/[\kappa^2(3N+1)^2+\Delta_C^2]$. Under the weak driving assumption, we have $N\ll M \ll 1$ and $M^2\approx N$, and $W\approx 0$ in the case of far off-resonance. As a result, the populations in one-photon and two-photon states are given by $p_{11}\approx N$ and $p_{22}\approx N^2$, respectively, which is the same as the results of thermal field driving~\cite{note1}. In the case of $\Delta_C=0$, we can obtain $W\approx -2M^2$ and the populations are given by $p_{11}\approx 0$ and $p_{22}\approx N/2$, which shows the unique feature of the squeezed vacuum injection. To verify these results, we numerically solve the Eq. (2) without any approximation. As shown in Fig. 2(b), the numerical results match the analytical ones very well.

\section*{Funding}
The National Key Basic Research Special Foundation (Grant No. 2016YFA0302800); the Shanghai Science and Technology Committee (STCSM) (Grants No. 18JC1410900); the National Nature Science Foundation (Grant No. 11774262).

\bibliography{references_bdq}

\begin{thebibliography}{10}
\bibitem{note1}
For weak driving fields (small $r$), we have $N=\sinh^2(r)\ll1$, which yielding
  $p_{11}=N$ and $p_{22}=N^2$ for the thermal field driving.
\end{thebibliography}






\end{document}